\documentclass[12pt,a4paper]{article}
\renewcommand{\baselinestretch}{1.2}
\input{epsf}
\begin{document}
%--------------------------------
%\documentclass[12pt]{article}
%\usepackage[cp1251]{inputenc}%%% russian in windows coding
%\usepackage[russian]{babel}

\oddsidemargin = -0.52cm     % отступ слева + 2,54 см
\textwidth = 16.96cm         % ширина содержимого (текста)
\topmargin = 0cm            % отступ сверху
\textheight = 23.7cm        % высота содержимого (текста)
\voffset = -0.54cm          % сдвиг страницы вверх
\paperwidth = 21cm          % ширина листа
\paperheight = 29.7cm       % высота листа
\renewcommand{\baselinestretch}{1.3}    % Интерлиньяж (интервал между строками:))
\parskip=6pt

\title{NoN analyticity  of  Hanle effect at zero magnetic field in a quantum dot}
\author{Kozlov G.G.}

\maketitle

\hskip 20mm Saint- Petersburg state university, spin optics laboratory

 \hskip20pt
%\vskip20pt
\hskip100pt {\it e}-mail:  gkozlov@photonics.phys.spbu.ru
%\vskip20pt
\begin{abstract}
Non analytic behaviour of Hanle effect in
 InGaAs quantum dots is described in terms of a simple 4-level model.
Despite simplicity the model  makes it possible to explain the observed fracture
 of Hanle curve at zero magnetic field  and obtain quantitative  agreement with the experiment.
\end{abstract}

\section{Introduction}

In this note we suggest a possible explanation of nonanalytic form of Hanle curves measured for an ensemble
of InGaAs/GaAs quantum dots (QD) in \cite{Cher}. These curves (i.e. the dependance of the degree of circular
polarization of luminescence on magnetic field in Voigt geometry) were obtained using the protocol of
excitation providing suppression of nuclear polarization in QD. Typical curve (taken from \cite{Cher})
presented at Fig.1a reveal a "fracture" at zero magnetic field.
High quality of Hanle curves obtained in \cite{Cher}
  make  this nonanalytic character unambiguous. Below we present an
explanation of this feature in the frame of a simple  model of energy structure of QD.
Similar model with some complications was used, for example, in \cite{Ign}.

\section{The model}

 This model consider QD as  an "island" of some semiconductor material
surrounded by a material with a broader band gap (in our case it is InGaAs QD surrounded by GaAs).
  Electron motion within  QD is quantized and
we  take into account only one level of spatial quantization
 in  valence zone (we call it {\it valence zone level}) and one level
in   conduction zone ({\it conduction zone level}).
  Each of these levels can be occupied by two electrons (with spin-up and
spin-down). We assume that the ground state $|0\rangle$ of  {\it uncharged } QD
 is correspond to the presence of two electrons at the valence zone level and
to the absence of electrons at the  conduction zone level.
Lowest  excitations of  QD can be obtained by  transition  of one electron to  the conduction zone level
 (appearance of hole at the valence level).
   So, four lowest excited states (equal to  the number of spin
configurations) are possible
   $|s_z,-\mu_z\rangle$, where
 $s_z=\pm 1/2$ and $\mu_z=\pm 1/2$  are
 spins of electron at the conducting and valence zone levels respectively.
 These four states we denote as
 $|++\rangle$, $|+-\rangle$, $|-+\rangle$, $|--\rangle$ (Fig.2).
Consider the possible channels of luminescence of  QD.
We assume that recombination of electron and hole with opposite spins is
 the only possible processes  giving rise to the luminescence
 and that both of these processes have the same probability. Therefore
\begin{equation}
\bigg|\langle ++|{\hat d}_{\pm}|0\rangle\bigg|^2=
\bigg|\langle --|{\hat d}_{\pm}|0\rangle\bigg|^2=0,
\end{equation}
and
\begin{equation}
\bigg|\langle +-|{\hat d}_+|0\rangle\bigg|^2=
 \bigg|\langle-+|{\hat d}_-|0\rangle\bigg|^2\equiv D^2,
\end{equation}
here ${\hat d}_{\pm}$ --
 is the dipole operator of interaction with electromagnetic
field of $\sigma_{\pm}$ polarization. The recombination of electron with spin +1/2 (-1/2)
 and hole with spin -1/2 (+1/2)  is accompanied
by  emission of $\sigma_+$ ($\sigma_-$) photon.
 Below we consider the rate of luminescence to be  proportional to $D^2$
\begin{equation}
W=k D^2
\end{equation}

\section{Luminescence and Hanle effect}

Within the frame of the  above model  an arbitrary excited state $\Psi$ of the QD can
be presented as a linear combination of the above basis states:
\begin{equation}
\Psi=C^{++}|++\rangle + C^{+-}|+-\rangle + C^{-+}|-+\rangle + C^{--}|--\rangle
\end{equation}

If $n$  QD's are prepared in such a state then within $dt$ time interval
 $dn_+$ photons in $\sigma_+$ polarization  will be emitted and
\begin{equation}
dn_+=
n k\bigg|\langle\Psi|d_+|0\rangle\bigg|^2dt=
 n W\bigg|C^{+-}\bigg|^2 dt
\label{5}
\end{equation}
 Analogously in $\sigma_-$
 polarization $dn_-$
 photons will be emitted
\begin{equation}
dn_-=
nk\bigg|\langle\Psi|d_-|0\rangle\bigg|^2dt
 =n W\bigg|C^{-+}\bigg|^2 dt
\label{6}
\end{equation}
If  emitted photons are registered  by a differential
polarimetric detector whose output current $i$
is proportional
to the difference of the photon currents in polarizations
$\sigma_+$ and $\sigma_-$ (i.e. $dn_+/dt - dn_-/dt$) with
corresponding coefficient of proportionality being the
quantum efficiency $\xi$ of the photoreceivers,
then the contribution  of  these $n$ QD's to the output current of such  detector can  be
calculated as
\begin{equation}
i=e\xi n W \bigg( \bigg|C^{+-}\bigg|^2 - \bigg|C^{-+}\bigg|^2\bigg)
\label{7}
\end{equation}
where $e$ is the electron charge.
In this formula the number $n$ of QD's and
 coefficients $C^{+-}$ and $C^{-+}$ can be time dependent.

Now we consider  the Hanle  effect i.e. the dependance of degree of polarization
 of luminescence on  magnetic field.
We imply the steadystate regime when the reasons exciting the luminescence do not depend on time.
For the process of QD's excitation we accept the following model. Assume
that at the time moment $T$
there are $N_0$
QD's in the ground state and there
is some action under which these dots (with probability per time unit
$P$ )
can be excited in  state
$\Psi_0$:
\begin{equation}
\Psi_0= C^{++}_0|++\rangle + C^{+-}_0|+-\rangle +
C^{-+}_0|-+\rangle + C^{--}_0|--\rangle
\label{8}
\end{equation}
with coefficients   $C^{++}_0$, $C^{+-}_0$, $C^{-+}_0$, $C^{--}_0$
 supposed  to be known.
(For example if this action  is irradiation
by light with $\sigma_+$ polarization then the only non zero coefficient is
 $C^{+-}_0$). So, in the case of this regime of excitation  $n_0=PN_0 dT$
 QD's in state $\Psi_0$ are generated within the time interval from $T$ to $T+dT$.
 Let us consider the contribution of this set of dots to
 the output current of the above differential receiver.
After excitation these  QD's begin to emit photons and their  contribution to the
output current  is described by Eq. (\ref{7}) in which
the temporary dependance of  coefficients $C^{+-}(t)$ and $C^{-+}(t)$
 is defined by   solution of  Shr\"odinger
equation for the wave function $\Psi(t)$ of QD with initial condition Eq. (\ref{8}) at $t=T$.
The relevant Hamiltonian has the form:
\begin{equation}
H=g_e\beta {\bf B s} +g_h\beta{\bf B \mu}
\end{equation}
where ${\bf s}$, ${\bf\mu}$ and ${\bf B}$ are the operators of electron and hole
(electron at the valence zone level) spins
 and magnetic field. In this formula we take into account that g-factors of electron $g_e$
and hole $g_h$ can be different: $g_e\ne g_h$.
 The temporary dependance of the number of excited QW's is governed by
the  equation which is obtained by summation of Eq (\ref{5}) and (\ref{6}):
\begin{equation}
{dn\over dt}= -n W\bigg(\bigg|C^{+-}(t)\bigg|^2+\bigg|C^{-+}(t)\bigg|^2\bigg)
\label{10}
\end{equation}
If $n(t=T)=n_0$ then one can write for  $n(t)$ the following expression:
\begin{equation}
n(t)=n_0\exp\bigg\{-W\int_T^t d\tau
\bigg(\bigg|C^{+-}(\tau)\bigg|^2+\bigg|C^{-+}(\tau)\bigg|^2\bigg)
\bigg\}
\end{equation}
Consequently, the contribution $di_T(t)$ of set of QD's created in  state
Eq. (8) in temporary interval $[T,T+dT]$ to the output current of differential
photodetector is defined by the following expression (see Eq. {\ref{7}}):
\begin{equation}
di_T(t)=\xi e P N_0 W dT \exp\bigg\{-W\int_T^t d\tau
\bigg(\bigg|C^{+-}(\tau)\bigg|^2+\bigg|C^{-+}(\tau)\bigg|^2\bigg)
\bigg\}\times
\end{equation}
$$
\times\bigg( \bigg|C^{+-}(t)\bigg|^2 - \bigg|C^{-+}(t)\bigg|^2\bigg)
\Theta(t-T)
$$
Assume that we have calculated the contribution $di_0(t)$ produced by QD's excited
at $T=0$. Due to the stationarity of $P$ and
$N_0$ (steadystate regime) the following relationship should hold:
\begin{equation}
di_T(t)=di_0(t-T)
\end{equation}
 To calculate the total output current $I$ of differential photodetector
 one should integrate contributions of all QD's excited at an arbitrary time moments:
\begin{equation}
I=\int_{-\infty}^{+\infty} {di_T(t)\over dT}\hskip2mm dT
\end{equation}
Taking into account the Eq's (12) and (13) and making the replacement of variable we obtain:
\begin{equation}
I=\xi e P N_0 W \int_0^{+\infty}dT \exp\bigg\{-W\int_0^T d\tau
\bigg(\bigg|C^{+-}(\tau)\bigg|^2+\bigg|C^{-+}(\tau)\bigg|^2\bigg)
\bigg\}\times
\end{equation}
$$
\times\bigg( \bigg|C^{+-}(T)\bigg|^2 - \bigg|C^{-+}(T)\bigg|^2\bigg)
$$
 Functions $C^{+-}(t)$ and $C^{+-}(t)$ entering this equation should be   obtained by solution of temporary
Shr\"odinger equation with initial conditions Eq. (8) at $t=0$.
Now we have to find the value of $N_0$ -- the number of  QD's in the ground state in steadystate
regime.  Let us for a moment consider the photoreceiver
 (not the above differential one) whose output current
 is proportional to the total photon current in both polarizations.
 If  we denote this current as $I_\Sigma$ then:
 $$
 I_\Sigma=\xi e P N_0
 $$
If the intensity $P$ of excitation is small and the effects of
saturation can be neglected then one can set the value of $N_0$
be equal to the total number $N$ of QD's in the region of irradiation.
 If it is not the case then for calculation  of $N_0$ the
total number of dots $N$ should be reduced by a number $N^\ast$
of dots excited to an arbitrary (because of the steadystate regime) time moment (say, $t$):
\begin{equation}
N_0=N-N^\ast
\end{equation}
 The value of $N^\ast$ can be obtained by summation of the
 numbers of QD's excited at all time moments $T$ with $T<t$
(see formula Eq. (11) ):
\begin{equation}
N^\ast=PN_0 \int_{-\infty}^t dT \exp\bigg\{-W\int_T^t d\tau
\bigg(\bigg|C^{+-}(\tau)\bigg|^2+\bigg|C^{-+}(\tau)\bigg|^2\bigg)
\bigg\}
\end{equation}
Using the fact that functions  $C(\tau)$ depend on the difference $\tau-T$
and making the relevant replacing of variables in the integral one can obtain:
\begin{equation}
N^\ast=PN_0 \int^{\infty}_0 dT \exp\bigg\{-W\int_0^T d\tau
\bigg(\bigg|C^{+-}(\tau)\bigg|^2+\bigg|C^{-+}(\tau)\bigg|^2\bigg)
\bigg\}
\end{equation}
 Substituting this into Eq. (16) we obtain for $N_0$ the following expression:
\begin{equation}
N_0=N\bigg[ 1+P \int^{\infty}_0 dT \exp\bigg\{-W\int_0^T d\tau
\bigg(\bigg|C^{+-}(\tau)\bigg|^2+\bigg|C^{-+}(\tau)\bigg|^2\bigg)
\bigg\}\bigg]^{-1}
\end{equation}
It is convenient  to introduce the following functions:
\begin{equation}
\Phi(T)\equiv\exp\bigg\{-W\int_0^T d\tau
\bigg(\bigg|C^{+-}(\tau)\bigg|^2+\bigg|C^{-+}(\tau)\bigg|^2\bigg),\hskip3mm
F(T)\equiv \bigg|C^{+-}(T)\bigg|^2 - \bigg|C^{-+}(T)\bigg|^2
\label{20}
\end{equation}
then
\begin{equation}
I=\xi e P N_0 W \int_0^{+\infty}dT \Phi(T)F(T)
\label{21}
\end{equation}
\begin{equation}
N_0=N\bigg[ 1+P \int^{\infty}_0 dT\Phi(T)\bigg]^{-1}
\label{22}
\end{equation}
To find the current $I$ one should calculate the temporary dependance of coefficients
 $C^{++}(t)$, $C^{+-}(t)$, $C^{-+}(t)$, $C^{--}(t)$.
We now solve this problem for the magnetic field having only $x$ component i.e.
 ${\bf B}=(B,0,0)$.
 We are going to analyse the Hanle effect in Voigt geometry, so this is the  only  case  of interest for us.
 If we arrange the basis functions corresponding to the presence of one
electron-hole pair in QD in the way described above then the matrix of
the Hamiltonian $H$ Eq. (9)  will take the following form:
\begin{equation}
H={\hbar\over 2}\left(\matrix{0 &  \nu_h &\nu_e&0\cr
\nu_h& 0 &0 & \nu_e \cr
 \nu_e&0&0&  \nu_h\cr
0&\nu_e&\nu_h &0}\right)
\end{equation}
where $\hbar\nu_{e(h)}\equiv g_{e(h)}\beta B$.
 All four states
in the absence of  magnetic field have the same energy  (degenerated).
This constant energy can be omitted in the
Shr\"odinger equation which has the form:
\begin{equation}
\imath {dC^{++}\over dt}=
{1\over 2}[\nu_h C^{+-}+\nu_e C^{-+}]
\label{24}
\end{equation}
$$
\imath {dC^{--}\over dt}=
{1\over 2}[\nu_e C^{+-}+\nu_h C^{-+}]
$$
$$
\imath {dC^{+-}\over dt}=
{1\over 2}[\nu_h C^{++}+\nu_e C^{--}]
$$
$$
\imath {dC^{-+}\over dt}=
{1\over 2}[\nu_e C^{++}+\nu_h C^{--}]
$$
Introduce new variables:
\begin{equation}
X\equiv C^{+-}+C^{-+},\hskip2mm
Y\equiv C^{++}+C^{--},\hskip2mm
Z\equiv C^{+-}-C^{-+},\hskip2mm
G\equiv C^{++}-C^{--}
\label{25}
\end{equation}
and
\begin{equation}
\Omega\equiv {1\over 2}{B\beta(g_h+g_e)\over \hbar}={1\over 2}(\nu_h+\nu_e),
 \hskip3mm
 \omega\equiv {1\over 2}{B\beta(g_h-g_e)\over \hbar}={1\over 2}(\nu_h-\nu_e)
\end{equation}
Then the general solution of Eq. (\ref{24}) has the form:
\begin{equation}
X=A\cos\Omega t + B\sin\Omega t,\hskip3mm
Y=\imath[B\cos\Omega t- A\sin\Omega t]
\end{equation}
\begin{equation}
Z=A_1\cos\omega t + B_1\sin\omega t,\hskip3mm
G=\imath[B_1\cos\omega t- A_1\sin\omega t],
\end{equation}
 where constants $A,A_1,B,B_1$ defined by the the initial conditions.
$C$- functions we are interesting in can be expressed as:
\begin{equation}
C^{+-}={1\over 2}(X+Z),\hskip10mm C^{-+}={1\over 2}(X-Z)
\end{equation}

\section{Excitation into the allowed for  the luminescence state }

In this case only  $C^{+-}=1$ is non-zero
 at $t=0$ and one can write down the following initial conditions:

\begin{equation}
X(t=0)=1,  \hskip3mm Y(t=0)=0,\hskip3mm Z(t=0)=1,\hskip3mm G(t=0)=0
\end{equation}
 Consequentely:
\begin{equation}
X(t)=\cos\Omega t, \hskip3mm Y(t)=-\imath\sin\Omega t, \hskip3mm
 Z(t)=\cos\omega t, \hskip3mm
G(t)=-\imath\sin\omega t
\end{equation}
and
\begin{equation}
C^{+-}={1\over 2}(\cos\Omega t+\cos\omega t), \hskip3mm
C^{-+}={1\over 2}(\cos\Omega t-\cos\omega t)
\end{equation}
For the combinations of $C$-functions we are interesting in we obtain:
\begin{equation}
\bigg|C^{+-}\bigg|^2+\bigg|C^{-+}\bigg|^2={1\over 2}(\cos^2\Omega t +\cos^2\omega t)
\end{equation}
and
\begin{equation}
\bigg|C^{+-}\bigg|^2-\bigg|C^{-+}\bigg|^2=\cos\Omega t\hskip2mm\cos\omega t
\end{equation}
In this case the functions Eq. (\ref{20}) (we supply them by a mark $b$ (bright))
 have the form :
\begin{equation}
\Phi_b(T)=\exp\bigg\{-{W\over 2}\bigg(
T+{\sin 2\Omega T\over 4\Omega}+{\sin 2\omega T\over 4\omega}
\bigg)\bigg\}
\label{35}
\end{equation}
$$
F_b(T)=\cos\Omega T\hskip2mm\cos\omega T
$$

Now it is possible to calculate the output current of differential
photodetector by means of formulas Eq. (22,21).
The field dependance (due to the field dependance of $\Omega$ and $\omega$)
 of the output current in this case is the curve
with wide maximum (nearly plane-like in the vicinity of zero field)
strongly differing from Lorentz curve (Fig3. top).

\section{ Excitation into the forbidden for the luminescence state }

This corresponds to the initial condition with only non-zero $C^{++}=1$.
Consequently (see Eq. (\ref{25}))
\begin{equation}
X(t=0)=0,  \hskip3mm Y(t=0)=1,\hskip3mm Z(t=0)=0,\hskip3mm G(t=0)=1
\end{equation}
 and then:
\begin{equation}
X(t)=-\imath\sin\Omega t, \hskip3mm Y(t)=\cos\Omega t, \hskip3mm
 Z(t)=-\imath\sin\omega t, \hskip3mm
G(t)=\cos\omega t
\end{equation}
and further:
\begin{equation}
C^{+-}=-{\imath\over 2}(\sin\Omega t+\sin\omega t), \hskip3mm
C^{-+}=-{\imath\over 2}(\sin\Omega t-\sin\omega t)
\end{equation}
For the combinations of $C$-functions we are interesting in we obtain:
\begin{equation}
\bigg|C^{+-}\bigg|^2+\bigg|C^{-+}\bigg|^2={1\over 2}(\sin^2\Omega t +\sin^2\omega t)
\end{equation}
and
\begin{equation}
\bigg|C^{+-}\bigg|^2-\bigg|C^{-+}\bigg|^2=\sin\Omega t\hskip2mm\sin\omega t
\end{equation}
In this case the functions Eq. (\ref{20}) (we supply them by a mark $d$ (dark))
 have the form:
\begin{equation}
\Phi_d(T)=\exp\bigg\{-{W\over 2}\bigg(
T-{\sin 2\Omega T\over 4\Omega}-{\sin 2\omega T\over 4\omega}
\bigg)\bigg\}
\label{41}
\end{equation}
$$
F_d(T)=\sin\Omega T\hskip2mm\sin\omega T
$$
Calculations of output current of differential photodetector
 by formulas Eq. (\ref{21}, \ref{22}) shows that current field dependance
 is non-analytic in the vicinity of zero field -- a kind of "fracture"  is appeared
 (Fig3. bottom). Note that the experimental Hanle curve also has the peculiarity of this type.

\section{ General case}

Let us consider now the general case.  It means that in the steadystate
regime there are $N_0$ dots in the ground state and $P^{+-}$
 is the probability of excitation of QD in state $|+-\rangle$ and $P^{++}$,  $P^{-+}$ and $P^{--}$
are the same probabilities for   states
 $|++\rangle$,  $|-+\rangle$ and   $|--\rangle$.
  In this case the calculation  analogous  to presented above give the
following expression for  output current of the differential photodetector in terms of
functions Eq. (\ref {35}) and Eq. (\ref{41})
 \begin{equation}
 I=\xi e W N_0\bigg[\bigg(P^{+-} - P^{-+}\bigg)\int_0^\infty dT F_b(T) \Phi_b(T)+
 \bigg(P^{++} - P^{--}\bigg)\int_0^\infty dT F_d(T) \Phi_d(T)\bigg]
 \end{equation}
$$
N_0=N\bigg[1+\bigg(P^{+-}+P^{-+}\bigg)\int_0^\infty \Phi_b(T)dT+
\bigg(P^{++}+P^{--}\bigg)\int_0^\infty \Phi_d(T)dT\bigg]^{-1}
$$
The commonly used quantity measured in experiment is  the degree of polarization (we denote it $\rho$).
 To calculate $\rho$ one should divide the obtained above differential photocurrent
 on total photocurrent  $I_\Sigma$ in both polarizations:
$$
  I_\Sigma=\xi e N_0(P^{+-}+P^{-+}+P^{++}+P^{--}),
$$
and consequently:
\begin{equation}
\rho=W{ \bigg[\bigg(P^{+-} - P^{-+}\bigg)\int_0^\infty dT F_b(T) \Phi_b(T)+
 \bigg(P^{++} - P^{--}\bigg)\int_0^\infty dT F_d(T) \Phi_d(T)\bigg]\over
P^{+-}+P^{-+}+P^{++}+P^{--}}
\label{43}
\end{equation}

\section{ Comparison with the experiment}

Now let us apply the obtained results to the experiments related to  Hanle  effect in InGaAs QD.
 First of all we note that the polarized luminescence is observed from the ensembles of {\it charged }
QD's \cite{Cher,Charged}. The degree of polarization of luminescence of
 uncharged QD was found to be weak. In charged QD
there are {\it three} electrons -- two electrons occupy the valence zone level and one the conducting
zone level.

In terms of above model the reason of weak polarization
 of the luminescence of uncharged QD's may be as follows. In uncharged QD
both electrons occupy the valence zone level. The external excitation generate the electron-hole pairs
{\it in the  barrier }. So, to emit photon the uncharged QD   should simultaneously trap  electron and hole from the
barrier. This may take much more time as compared with that required for trapping {\it only}
electron (or only hole). Therefore the initial polarization of the electron-hole pair (created by
the external polarized pumping) can decay and  the luminescence of QD appears to be weakly polarized in
this case.

Now let QD be charged by a single electron (resident electron) as it is the case for the experiments
described in\cite{Cher}. In this case of {\it negatively} charged QD  trapping of the hole
 ({\it positively charged}) from the barrier is
likely due to the Coloumn attraction.  For this reason  trapping of the hole can be so fast that its polarization
does not decay and is defined by the polarization of pumping
 (say $\sigma_+$).
   Therefore if the spin of resident electron is random then we have the case of excitation of   QD states
$|+->$ and $|-->$ with equal probabilities. The scheme of this process is presented at fig.4.
 So, to calculate the Hanle signal one can use formula
(\ref{43}) with $P^{--}=P^{+-}\ne 0$ and $P^{++}=P^{-+}=0$. The result of such calculation is presented at
fig.1b. The values of fitting parameters are presented at figure caption. One can see that theoretical
and experimental curves have much in common. This justify the calculation presented in this note.
 Despite the fact that the only important fitting parameter in our
 calculations was the ratio $g_e/g_h$ the degree of polarization at zero field  ($17 \%$)
 is also in good agreement with the experiment.

The most important feature of the observed Hanle curve is, in our opinion, the presence of a fracture in
the vicinity of zero field. Within the above simple model this fracture appeared due to the excitation of
QD in the state $|++>$ forbidden for the luminescence  (see section V).
The second important reason of this is temporal dependance of the
probability of the luminescence after excitation (see Eq. (\ref{10})).
Similar dependance was studied, for example, in \cite{Dyakonov}

Finally it should be mentioned that one of the curious  features of the
luminescence of QD is that its polarization has the  sign  inverse to that
of excitation \cite{Cher,Ign}. We are not ready to discuss this phenomenon in details
but it  sounds like true  that the hole can change  spin  polarization  during
 trapping by the QD.

The author thank Cherbunin R.V., Yugova I.A., Ignatiev I.V., Gerlovin I. Ya. for discussions.

Sorry for not perfect English.

\newpage

\begin{figure}
\epsfxsize=400pt \epsffile{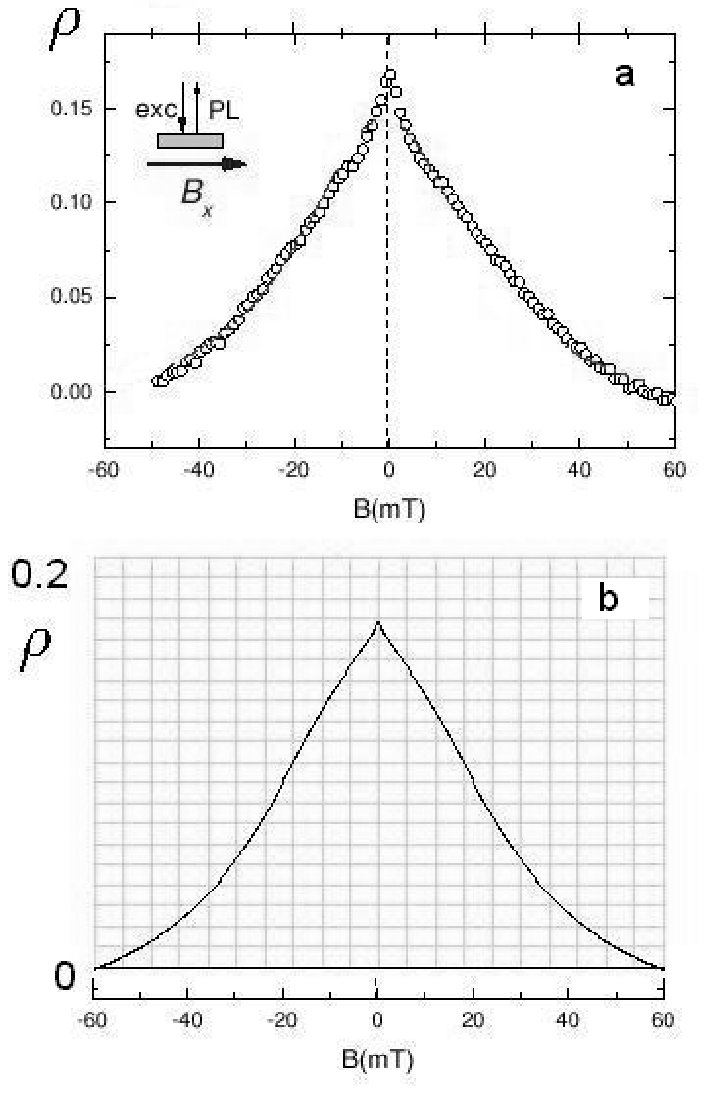} \caption{(a) -- the experimental Hanle curve for InGaAs QD ensemble,
(b) -- the same calculated.  The fitting parameters are: $ g_h=0.28, g_e=0.57,
 P_{--}=1, P_{+-}=1, P_{--}=P_{-+}=0, W=0.25 \cdot 10^{10}$ sec$^{-1}$.}
\end{figure}

\begin{figure}
\epsfxsize=400pt \epsffile{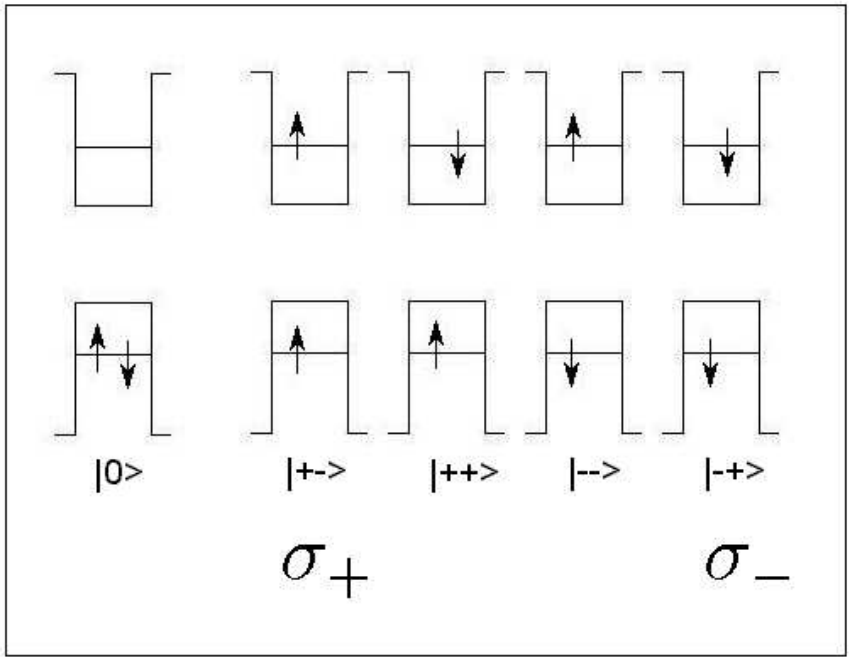} \caption{The QD's ground and excited states.}
\end{figure}

\begin{figure}
\epsfxsize=400pt \epsffile{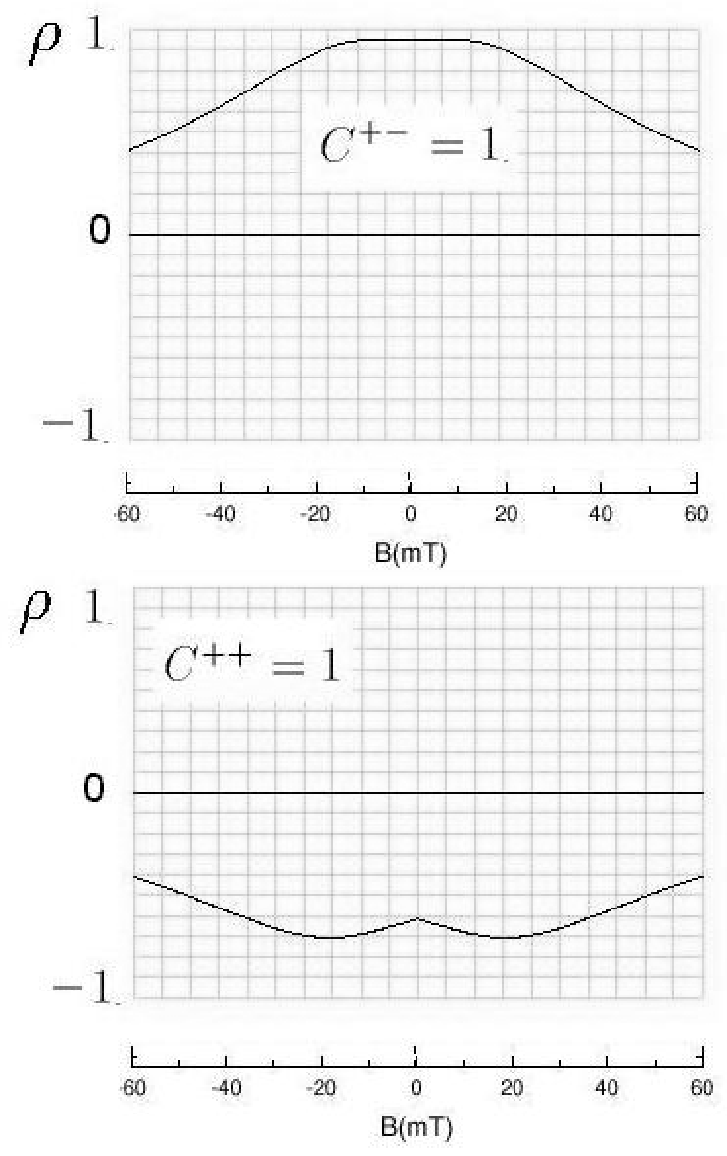} \caption{Top -- the Hanle curve for the case of excitation in $|+->$ state,
 bottom -- the same for the case of excitation in $|++>$ state.}
\end{figure}

\begin{figure}
\epsfxsize=400pt \epsffile{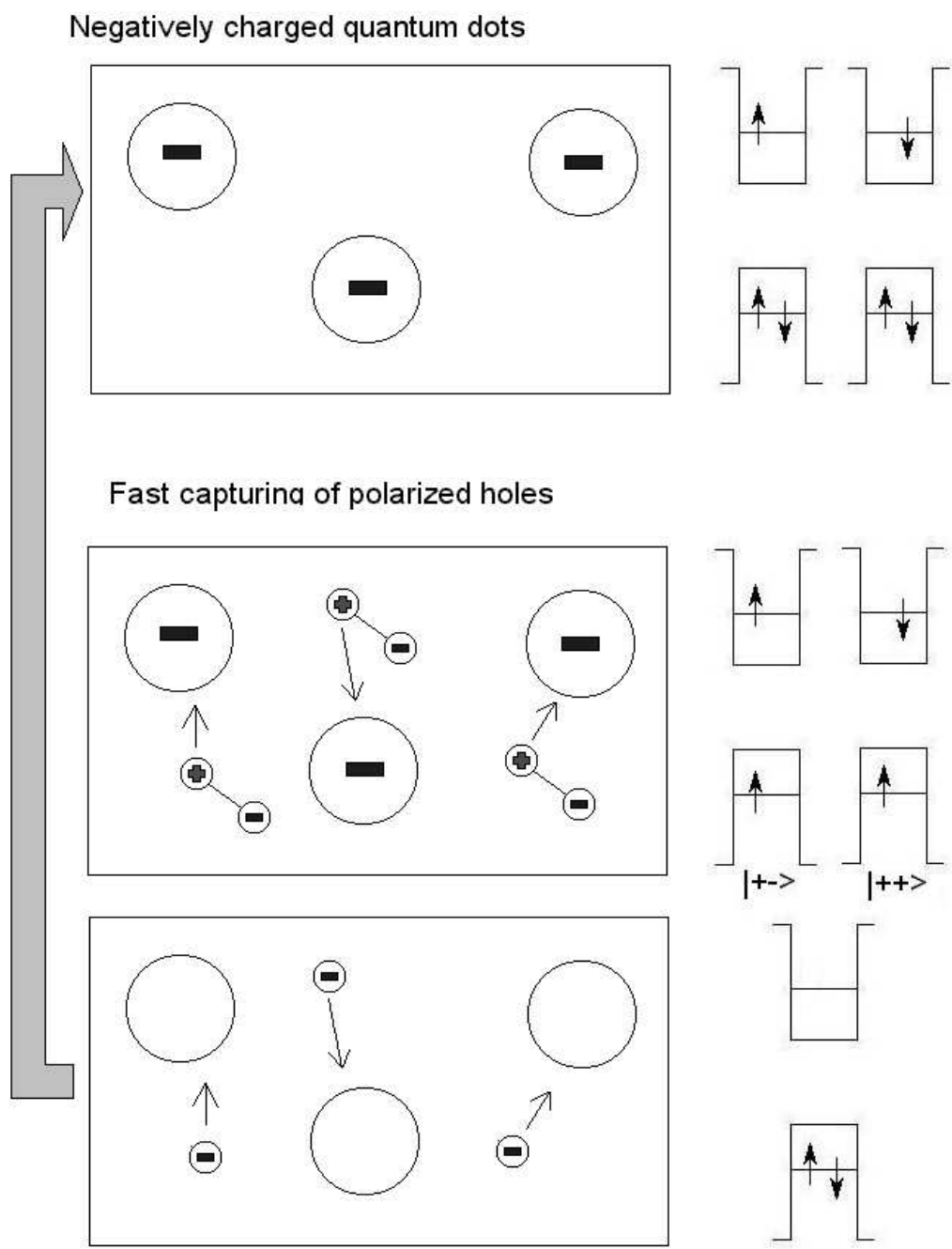} \caption{Scheme of QD excitation.}
\end{figure}

\end{document}